\documentclass[letterpaper, 10 pt, conference]{ieeeconf}  

\IEEEoverridecommandlockouts                              

\usepackage[utf8]{inputenc}
\usepackage{blindtext}
\usepackage{graphicx}
\usepackage[utf8]{inputenc}
\usepackage{graphics} 
\usepackage{epsfig} 
\usepackage{mathptmx} 
\usepackage{amsmath} 
\usepackage{amssymb}  
\usepackage{cite}
\usepackage{multicol}
\usepackage{mathtools, cuted}
\usepackage[cmintegrals]{newtxmath}
\usepackage{epstopdf}
\usepackage{graphics} 
\usepackage{epsfig} 

 \usepackage{subfigure}
\usepackage{titlesec}
\usepackage{epsfig} 
\usepackage{mathptmx} 
\usepackage{amsmath} 
\usepackage{amssymb}  
\usepackage{cite}
\usepackage{multicol}
\usepackage{mathtools, cuted}
\usepackage[cmintegrals]{newtxmath}
\usepackage{makeidx}         
\usepackage{algorithmic}

\usepackage{multicol}        
\usepackage[bottom]{footmisc}
\usepackage{caption}
\usepackage{nomencl}
\usepackage[usenames, dvipsnames]{color}

 \usepackage{amsthm}
 \theoremstyle{definition}

\theoremstyle{theorem}
\newtheorem{theorem}{Theorem}

\theoremstyle{assumption}
\newtheorem{assumption}{Assumption}

\theoremstyle{lemma}

\theoremstyle{remark}

 \theoremstyle{proposition}

\usepackage{algorithmic}    

\usepackage{array}
\usepackage[ruled,vlined]{algorithm2e}

\title{\LARGE \bf
Conservation-Based Modeling and Boundary Control of Congestion with an Application to Traffic Management in Center City Philadelphia}

\author{Xun Liu and Hossein Rastgoftar$^{2}$
\thanks{The authors are with the Department of Mechanical Engineering at Villanova University, Villanova, PA 19085, USA {\tt\small \{xliu8, hossein.rastgoftar\}@villanova.edu}}%
}

\begin{document}

\maketitle
\thispagestyle{empty}
\pagestyle{empty}

\begin{abstract}

This paper develops a conservation-based approach to model traffic dynamics and alleviate traffic congestion in a network of interconnected roads (NOIR). We generate a NOIR by using the Simulation of Urban Mobility (SUMO) software based on the real street map of Philadelphia Center City. The NOIR is then represented by a directed graph with nodes identifying distinct streets in the Center City area. By classifying the streets as inlets, outlets, and interior nodes, the model predictive control (MPC) method is applied to alleviate the network traffic congestion by optimizing the traffic inflow and outflow across the boundary of the NOIR with consideration of the inner traffic dynamics as a stochastic process. The proposed boundary control problem is defined as a quadratic programming problem with constraints imposing the feasibility of traffic coordination, and a cost function defined based on the traffic density across the NOIR.

\end{abstract}

\section{Introduction}

In the process of urbanization and the rapid popularization of private vehicles, the problem of urban traffic congestion has become more and more prominent and produced numerous negative impacts on economy \cite{muneera2018economic, margaret2004impact} and environment  \cite{liang2013road, annan2015traffic}. Traffic congestion can destroy the urban environment and ecology. Due to the low-speed driving conditions, the emission of greenhouse gas, noxious gas and noise will increase, and that will badly affect human health \cite{chin2011impact}. 

Over the past decades, a large number of scholars have developed prediction, control, and optimization methods to solve the challenges of traffic congestion in urban areas. Ref.\cite{xiao2014hierarchical} offers an integration of fuzzy rule-based systems and the genetic algorithms  to model and predict the traffic coordination. Refs.\cite{toshio2017predicting} and \cite{zhao2019peak} develop the traffic predictive approaches by relying on driver  behavior and bus driving intervals. With the rapid development of V2X and autonomous driving technology, floating car data (FCD) technology has been widely used to estimate the traffic state \cite{xiangjiekong2016urban, tettamanti2017nonlinear}. 

Researchers have also developed different model-based and model-free approaches to obtain dynamics of traffic coordination and control congestion. The model-based macroscopic fundamental diagram (MFD), whose applicability for urban traffic is experimentally verified in \cite{geroliminis2008existence}, is an efficient tool to obtain dynamics of an urban traffic network. Ref.\cite{xu2013traffic} applies the MFD model to evaluate the traffic accumulation amount, and estimate the traffic state. Refs. \cite{sirmatel2017integration, li2019model-free} integrate MFD with perimeter control to improve the mobility of a traffic network. Moreover, Refs.\cite{yang2017fundamental,shao2018distributed,munoz2003traffic,yin2017offblock,feldman2002cell} apply the cell transmission model (CTM) method to enhance the efficiency and accuracy of the network modeling by partitioning the traffic network into homogeneous road elements. Recently, with the improvement of computing capacity and the development of AI technology, reinforcement learning (RL) method has attracted more and more attention. Ref. \cite{greguric2020application} presents an overview of the recently-developed RL algorithms in the area of adaptive traffic signal control. In Refs.\cite{lin2018efficient,abdulhai2003reinforcementlearning,mannion2016anexperimental, prashanth2011reinforcementlearning}, researchers integrate the model-free methods with RL approaches to optimally plan the functionalities of traffic signals. The model predictive control (MPC) is another commonly used tool for controlling the traffic dynamics in urban networks. Refs. \cite{li2019model-free}, \cite{sirmatel2017integration}, \cite{rastgoftar2019integrative}, \cite{rastgoftar2020resilient} and \cite{rastgoftar2021physicsbased} apply the MPC approach to assign optimal boundary control variables. Ref.\cite{lin2011fast} integrates the MPC and mixed-integer linear programming (MILP) to manage the complexity of traffic coordination optimization.

\begin{figure}[!ht]
    \centering
    \includegraphics[width=0.48\textwidth]{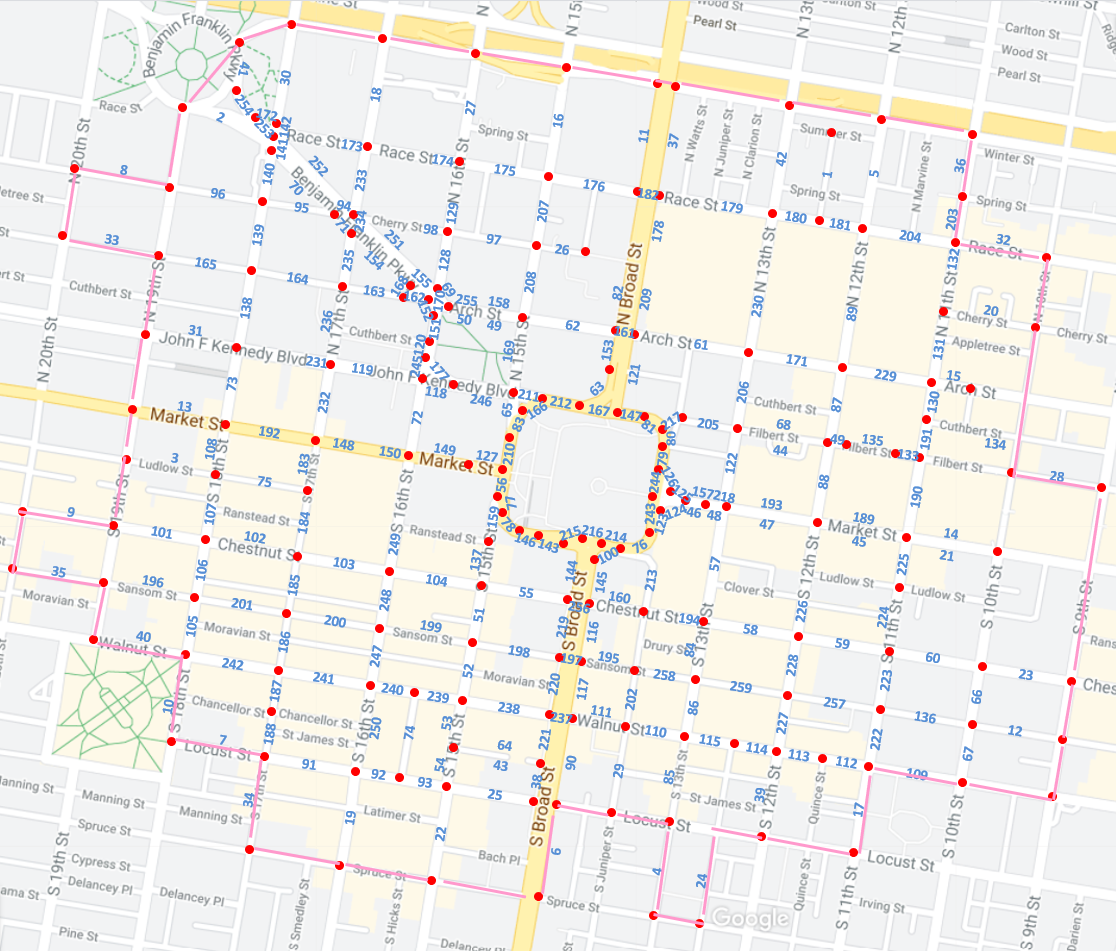}
    \caption{Example NOIR: Center City, Philadelphia}
    \label{fig:Philadelphia Center}
    \vspace{-1.3em}
\end{figure}

This paper offers an integration of mass conservation law and MPC-based boundary control to obtain the traffic dynamics and alleviate the traffic congestion. We use the Simulation of Urban Mobility (SUMO) software to convert the real street map data into a directed graph representing a network of inter-connected roads (NOIR). While we previously modeled traffic inner dynamics as a time-invariant stochastic process in Refs.  \cite{rastgoftar2019integrative, rastgoftar2020resilient, rastgoftar2021physicsbased}, this paper applies
the mass conservation law to model the traffic inner dynamics as a non-stationary stochastic process and obtain the traffic feasibility conditions at the interior nodes. Compared to Refs. \cite{rastgoftar2019integrative, rastgoftar2020resilient, rastgoftar2021physicsbased} that control the congestion only through the inlet boundary nodes, we apply the MPC to control the boundary inflow through the NOIR inlets, and the boundary outflow through the NOIR outlets. For the case study, we use the proposed model and control approach to evaluate congestion management in a certain area of Center City Philadelphia with the map shown in Fig. \ref{fig:Philadelphia Center}. 

This paper is arranged in the following structure: Section \ref{section 2} explains the preliminary notions of graph theory. The problem statement is presented in Section \ref{section 3} and followed by obtaining the traffic network dynamics and providing the feasibility conditions in Section \ref{section 4}. Section \ref{section 5} presents the boundary control approach based on the MPC method to control the traffic congestion. Then, the simulation results are reported in Section \ref{section 6} and followed by the conclusion in Section \ref{section 7}.

\section{Preliminary Notions of Graph Theory}\label{section 2}
In this paper, an NOIR is represented by graph $\mathcal{G}\left(\mathcal{V},\mathcal{E}\right)$ where $\mathcal{V}$ and $\mathcal{E}\subset \mathcal{V}\times \mathcal{V}$ define nodes and edges of graph $\mathcal{G}$, respectively. We use $i\in\mathcal{V}$ to represent a road element in the NOIR. Note that all of the road elements partitioned and generated by SUMO in the NOIR are unidirectional. The bidirectional
road is presented by two parallel one-way road elements (See Fig.\ref{fig:Philadelphia Center}). Also, since we are more interested in the boundary traffic dynamics, the traffic light controller is not in consideration in this paper. Edge $\left(i,j\right)\in \mathcal{E}$ represents a directed connection from road element $i\in \mathcal{V}$ to road element $j\in \mathcal{V}$.

Set $\mathcal{V}$ can be partitioned as $\mathcal{V}=\mathcal{V}_{in}\bigcup \mathcal{V}_{out}\bigcup \mathcal{V}_{I}$, where subsets $\mathcal{V}_{in}=\{1,\cdots,N_{in}\}$, $\mathcal{V}_{out}=\{N_{in}+1,\cdots,N_{out}\}$, and $\mathcal{V}_{I}=\{N_{out}+1,\cdots,N\}$ define the index numbers of inlets, outlets, and interior road elements respectively. For every road element $i\in \mathcal{V}$, sets
\begin{subequations}
    \begin{align}
        \mathcal{I}_i&=\left\{j:\left(j,i\right)\in \mathcal{E}\right\},\\
        \mathcal{O}_i&=\left\{j:\left(i,j\right)\in \mathcal{E}\right\}
    \end{align}
\end{subequations}
define in-neighbors and out-neighbors. Traffic is directed from in-neighbor $j\in \mathcal{I}_i$ towards $i\in \mathcal{V}\setminus \mathcal{V}_{in}$, or it is directed from $i\in \mathcal{V}\setminus \mathcal{V}_{out}$ towards out-neighbor $j\in \mathcal{O}_i$.

\section{Problem Statement}\label{section 3}

We implement the mass-conservation law to obtain traffic dynamics at every road element $i\in\mathcal{V}$. Let $s_i[k]$ denote the external flow, and $\rho_{i}[k]$, $y_i[k]$, and $z_i[k]$ denote traffic density, network inflow, and network outflow of road $i\in \mathcal{V}$, respectively. Then, traffic dynamics at road element $i\in \mathcal{V}$ can be defined by
\begin{equation}\label{Equation: 1}
    \rho_{i}[k+1]=\rho_{i}[k]+y_{i}[k]-z_{i}[k]+s_{i}[k]
\end{equation}
where $k=0,1,2,\cdots$ denotes the discrete sampling time.

External flow $s_{i}[k]$ quantifies the traffic inflow entering the NOIR through inlet road element $i\in \mathcal{V}_{in}$, or the traffic outflow leaving the NOIR through outlet road element $i\in \mathcal{V}_{out}$ within time interval $\left[t_{k},t_{k+1}\right]$. We define $s_i[k]$ as follows:
\begin{equation}\label{external flow}
    s_{i}[k]=
    \begin{cases}
    u_{i}[k]\geq 0   & {i\in \mathcal{V}_{in}}\\
    -v_{i}[k]\leq 0  & {i\in \mathcal{V}_{out}}\\
    0                & {i\in \mathcal{V}_I}
    \end{cases}
    .
\end{equation}

Network inflow $y_{i}[k]$ and network outflow $z_{i}[k]$ are given by
\begin{subequations}\label{network inflow and outflow}
\newlength{\widest}
\settowidth{\widest}{$v_{i}[k]=q_{i,j}[k]z_{j}[k]$}
\begin{align}
    y_{i}[k]=&
    \begin{cases}\label{network inflow}
    0&\quad i\in\mathcal{V}_{in}\\
    v_{i}[k]&\quad i\in\mathcal{V}_{out}\\
    \sum_{j\in \mathcal{I}_i}q_{i,j}[k]z_{j}[k]&\quad i\in\mathcal{V}_{I}
    \end{cases},\\
    z_{i}[k]=&
    \begin{cases}\label{network outflow}
    \makebox[\widest][l]{$u_{i}[k]$}&\quad i\in\mathcal{V}_{in}\\
    0&\quad i\in\mathcal{V}_{out}\\
    p_{i}[k]\rho_{i}[k]&\quad i\in\mathcal{V}_{I}
    \end{cases}
    ,
\end{align}
\end{subequations}
where
\begin{equation}\label{outflowprobability}
    p_i[k]=\begin{cases}
    1&\mathrm{If~}i\in \mathcal{V}_{in}\bigcup\mathcal{V}_{out}\\
    0&\mathrm{If~}i\in \mathcal{V}_{I}~\mathrm{and}~\rho_i=0\\
    \frac{z_i[k]}{\rho_i[k]}&\mathrm{If~}i\in \mathcal{V}_{I}~\mathrm{and}~\rho_i\neq0\\
    \end{cases}
\end{equation}
is the outflow probability of road element $i\in\mathcal{V}_{I}$ at discrete time $k$, $q_{i,j}\in\left[0,1\right]$ is the fraction of outflow traffic directed from $j\in \mathcal{V}\setminus\mathcal{V}_{out}$ to $i\in O_{j}$ at every discrete time $k$, and
\begin{equation}
    \sum_{i\in O_{j}} q_{i,j}= 1
\end{equation}
at every interior road $i\in \mathcal{V}_I$.

Given the above problem setting, the main purpose of this paper is to alleviate the traffic congestion by assigning the optimal external flow $s_{i}[k]$. Assuming $p_i[k]$ and $q_{i,j}[k]$ are known at every interior road $i\in \mathcal{V}_I$, the external flow is determined by solving a quadratic programming problem with cost function 
\begin{equation}\label{cost function}
  \mathrm{C}=
  {\frac{1}{2}}\sum_{l=0}^{N_\tau-1}\left(\sum_{i\in \mathcal{V}_{in}}u_i^2[k+l]+\sum_{j\in \mathcal{V}_{out}}v_j^2\left[k+l\right]\right)
\end{equation}
and the following inequality and equality constraints:
\begin{subequations}\label{constrain condition}
\begin{gather}
     \bigwedge_{l=0}^{N_\tau-1}\bigwedge_{i\in \mathcal{V}_{in}}\left(u_i\left[k+l\right]\geq 0\right), \label{c1}\\
     \bigwedge_{l=0}^{N_\tau-1}\bigwedge_{j\in \mathcal{V}_{out}}\left(v_j\left[k+l\right]\geq 0\right), \label{c2}\\
     \bigwedge_{l=0}^{N_\tau-1}\bigwedge_{i\in \mathcal{V}_{I}}\left(\rho_i\left[k+l\right]\geq 0\right), \label{c3}\\
     \bigwedge_{l=0}^{N_\tau-1}\bigwedge_{i\in \mathcal{V}_{I}}\left(\rho_i\left[k+l\right]\leq \rho_{i,max}\right), \label{c4}\\
     \bigwedge_{l=0}^{N_\tau-1}\left(\sum_{i\in \mathcal{V}_{in}}u_i[k+l]+\sum_{j\in \mathcal{V}_{out}}v_j\left[k+l\right]=d_0\right) \label{c5}.
\end{gather}
\end{subequations}
Note that $N_\tau<\infty$ is the time horizon length and $\rho_{i,max}$ is the maximum number of vehicles that can be accommodated at road element $i\in \mathcal{V}_I$. Constraint Eqs. \eqref{c1} and \eqref{c2} ensure that the traffic back-flow is avoided at every inlet or outlet road element. Constraint \eqref{c3} ensures that the solution of the above optimization problem obtains a non-negative traffic density distribution across the NOIR. Constraint \eqref{c4} is imposed to avoid the traffic over-saturation. Assuming the demand for entering and leaving the NOIR is sufficiently high, equality constraint \eqref{c5} ensures that $d_0$ cars can cross the border of the NOIR at every discrete time $k$.

\section{Traffic Network Dynamics}\label{section 4}

By substituting \eqref{external flow} and \eqref{network inflow and outflow} in \eqref{Equation: 1}, the traffic dynamics at road element $i\in \mathcal{V}$ simplifies to
\begin{subequations}\label{traffic dynamics}
\begin{gather}
    \forall i\in\mathcal{V}_{in}\bigcup\mathcal{V}_{out},\quad\rho_i[k+1]=\rho_i[k] \label{traffic dynamics boundary node}\\
\resizebox{0.99\hsize}{!}{%
    $
    \forall i\in\mathcal{V}_{I},\quad\rho_i[k+1]=\left(1-p_i[k]\right)\rho_i[k]+\sum_{j\in \mathcal{I}_i}q_{i,j}[k]p_j[k]\rho_j[k].
    $
    } \label{traffic dynamics interior node}
\end{gather}
\end{subequations}

Eq. \eqref{traffic dynamics boundary node} shows that the traffic density remains constant at inlet and outlet road elements. Thus, traffic dynamics are only defined for the interior road elements. 
\begin{assumption}\label{assumption1}
Discrete time $k$ represents the time interval $\left[t_{k},t_{k+1}\right)$ where the time increment $\Delta T=t_{k+1}-t_k$ is assumed to be constant for  $k=0,1,2,\cdots$. We choose a sufficiently-small time increment $\Delta T$ such that
\begin{equation}
    p_i[k]=\dfrac{z_i[k]}{\rho_i[k]}\in \left[0,1\right)
\end{equation}
at every road element $i\in \mathcal{V}_I$ and every discrete time $k$, if $\rho_i[k]\neq 0$. Note that $p_i[k]=0$, if $\rho_i[k]=0$ over the time interval $t\in \left[t_k,t_{k+1}\right)$ (See Eq. \eqref{outflowprobability}).
\end{assumption}
To obtain the traffic network dynamics, we define state vector $\mathbf{x}=[\rho_{N_{out}+1} \cdots \rho_N]^\mathsf{T}\in\mathbb{R}^{\left(N-N_{out}\right)\times 1}$, NOIR inflow vector $\mathbf{y}=[y_{N_{out}+1} \cdots y_N]^\mathsf{T}\in \mathbb{R}^{\left(N-N_{out}\right)\times 1}$, 
NOIR outflow vector $\mathbf{z}=[
z_{N_{out}+1} \cdots z_N]^\mathsf{T}\in \mathbb{R}^{\left(N-N_{out}\right)\times 1}$, outflow probability matrix $\mathbf{P}=\mathrm{diag}\left(p_{N_{out}+1},\cdots,p_N\right)\in \mathbb{R}^{\left(N-N_{out}\right)\times \left(N-N_{out}\right)}$, and tendency probability matrix $\mathbf{Q}=\left[Q_{ij}\right]$ with $(i,j)$ entry
\begin{equation}
    Q_{ij}[k]=q_{i+N_{out},j+N_{out}}[k].
\end{equation}
Note that $q_{i+N_{out},j+N_{out}}[k]$ is the fraction of outflow of road $\left(j+N_{out}\right)\in \mathcal{V}_I$ directed towards $\left(i+N_{out}\right)\in \mathcal{V}_I$.

Per Eq. \eqref{network inflow}, the NOIR inflow vector $\mathbf{y}$ and outflow vector $\mathbf{z}$ can be related to $\mathbf{x}$ by
\begin{subequations}
\begin{align}
    \mathbf{y}[k]&=\mathbf{Q}[k]\mathbf{P}[k]\mathbf{x}[k],\label{y15}\\
    \mathbf{z}[k]&=\mathbf{P}[k]\mathbf{x}[k]. \label{rawz}
\end{align}
\end{subequations}

If Eq. \eqref{traffic dynamics interior node} is applied to model traffic coordination at every interior node $i\in \mathcal{V}_I$, the traffic network dynamics become
\begin{equation}\label{traffic dynamic state vector}
    \mathbf{x}[k+1]=\mathbf{A}[k]\mathbf{x}[k]+\mathbf{B}[k]\mathbf{s}[k],
\end{equation}
where $\mathbf{s}[k]=\left[s_{i}[k]\right]\in\mathbb{R}^{N_{out}\times1}$, $\mathbf{B}[k]=[b_{ij}[k]]\in \mathbb{R}^{\left(N-N_{out}\right)\times N_{out}}$, $\mathbf{A}[k]\in \mathbb{R}^{\left(N-N_{out}\right)\times \left(N-N_{out}\right)}$ are obtained as follows:
\begin{subequations}
\begin{gather}
    s_{i}[k]=
    \begin{cases}
    u_{i}[k],&\mathrm{If~} i\in\mathcal{V}_{in}=\{1,\cdots,N_{in}\}\\
    v_{i}[k],&\mathrm{If~} i\in\mathcal{V}_{out}=\{N_{in}+1,\cdots,N_{out}\}
    \end{cases},\\
    b_{ij}[k]=
    \begin{cases}
    1\quad j\in\mathcal{I}_{i+N_{out}}\\
    -1\quad j\in\mathcal{O}_{i+N_{out}}
    \end{cases},\\
    \mathbf{A}[k]=\mathbf{I}-\mathbf{P}[k]+\mathbf{Q}[k]\mathbf{P}[k].
\end{gather}
\end{subequations}
\begin{theorem}\label{theorem1}
Assume graph $\mathcal{G}\left(\mathcal{V},\mathcal{E}\right)$ holds the following properties:
\begin{enumerate}
    \item{Traffic inflow directs from every inlet boundary road element towards an interior road element.}
    \item{There is at least one directed path from every interior node to an outlet node.}
    \item{Graph $\mathcal{G}$ contains no isolated node.}
    \item{No inlet boundary road element is directly connected to an outlet boundary road element.}
\end{enumerate}
Then, the traffic network dynamics \eqref{traffic dynamic state vector} is BIBO stable.
\end{theorem}
\textbf{Proof:} 
If assumptions of Theorem \ref{theorem1} are satisfied, matrix $\mathbf{A}[k]$ holds the following properties at every discrete time $k$:
\begin{enumerate}
    \item{All entries in matrix $\mathbf{A}[k]$ are non-negative.}
    \item{Column $i$ of matrix $\mathbf{A}[k]$ sums up to $1$, if $\mathcal{O}_{i+N_{out}}\bigcap \mathcal{V}_{out}=\emptyset$.}
    \item{Elements of column $i$ of matrix $\mathbf{A}[k]$ sums up to a positive number in interval $(0,1)$, if $\mathcal{O}_{i+N_{out}}\bigcap \mathcal{V}_{out}\neq\emptyset$.}
\end{enumerate}
Thus, eigenvalues of matrix $\mathbf{A}[k]$ are all less than one at every discrete time $k$.

Per traffic dynamics \eqref{traffic dynamic state vector}, we can define
\begin{equation}
\label{MAINNNNTheorem4}
\begin{split}
 \mathbf{x}[k+1]=&
 \mathbf{\Theta}_k
 \begin{bmatrix}
    \mathbf{x}[1]\\
    \mathbf{B}[k]\mathbf{s}[1]\\
    \vdots\\
    \mathbf{B}[k]\mathbf{s}[k]\\
    \end{bmatrix}
    ,
   \end{split}
\end{equation}
where 
\begin{subequations}
\begin{align}
\mathbf{\Theta}_k&=
\begin{bmatrix}
 \mathbf{\Gamma}_{k}&\cdots&\mathbf{\Gamma}_1&\mathbf{\Gamma}_0
\end{bmatrix},\\
\mathbf{\Gamma}_{h}&=\prod_{j=k-h+1}^k \mathbf{A}[j],
\end{align}
\end{subequations}
for $h=1,\cdots,k$, and $\mathbf{\Gamma}_0=\mathbf{I}_{N-N_{out}}\in \mathbb{R}^{\left(N-N_{out}\right)\times \left(N-N_{out}\right)}$ is an identity matrix. Because $\mathbf{x}[1]<\infty$, and $\mathbf{s}[k]$ is bounded at every discrete time $k$, we can write
\begin{subequations}
\begin{align}
    \mathbf{x}[1]\leq z_{max}\mathbf{1}_{N-N_{out}\times1},\\
    \mathbf{B}[k]\mathbf{s}[k]\leq z_{max}\mathbf{1}_{N-N_{out}\times1},
\end{align}
\end{subequations}
where $z_{\mathrm{max}}$ is bounded. If assumptions of Theorem \ref{theorem1} are satisfied, spectral radius $r$ of matrix $\mathbf{\Gamma}_k$ is less than 1 at every discrete time $k$. Therefore, we can write
\begin{equation}\label{boundedproof_spe}
\resizebox{0.99\hsize}{!}{%
    $
\begin{split}
    \mathbf{x}^\mathsf{T}\left[k+1\right]\mathbf{x}\left[k+1\right]&\leq z_{max}\mathbf{1}_{N-N_{out}\times1}^\mathsf{T}\left(\sum_{l=0}^k\sum_{h=0}^k\mathbf{\Gamma}_l^\mathsf{T}\mathbf{\Gamma}_h\right)z_{max}\mathbf{1}_{N-N_{out}\times1}\\
    &\leq z_{max}^2\left(N-N_{out}\right)\left(\sum_{l=0}^\infty r^l\right)\leq\dfrac{z_{max}^2\left(N-N_{out}\right)}{\left(1-r\right)}
\end{split}
$
}
\end{equation}
which implies that $\mathbf{x}^\mathsf{T}\left[k+1\right]\mathbf{x}\left[k+1\right]$ is bounded  at every discrete time $k$, and thus the BIBO stability of traffic dynamics \eqref{traffic dynamic state vector} is proven.

\section{Traffic Congestion Control}\label{section 5}

This paper applies the model predictive control (MPC) approach to control congestion through optimizing the boundary inflow and outflow. For the proposed MPC control, we use the linear time-varying dynamics \eqref{traffic dynamic state vector} to predict traffic evolution within a future finite time horizon, and determine the optimal boundary external flow as a solution of the quadratic function subject to the inequality and equality constrains.

Given traffic dynamics \eqref{traffic dynamic state vector} at discrete time, the following predictive model can be used to model traffic coordination within the next $N_\tau$ time steps:
\begin{equation}\label{iterative model}
    \mathbf{X}[k]=\mathbf{G}[k]\mathbf{x}[k]+\mathbf{H}[k]\mathbf{U}[k]
\end{equation}
where
\begin{subequations}\label{state iternative model}
\begin{gather}
    \mathbf{X}[k]=\begin{bmatrix}
    \mathbf{x}^\mathsf{T}[k+1]&
    \cdots&
    \mathbf{x}^\mathsf{T}[k+N_\tau]
    \end{bmatrix}^\mathsf{T}
    \in \mathbb{R}^{\left(N_\tau N\right)\times 1}, \label{Vector X}\\
    \mathbf{G}[k]=\begin{bmatrix}
    \mathbf{A}[k]\\
    \vdots\\
    \mathbf{A}^{N_\tau}[k]
    \end{bmatrix}
    \in \mathbb{R}^{\left(N_\tau N\right)\times N}, \label{MatrixG}\\
    \resizebox{0.99\hsize}{!}{%
    $
    \mathbf{H}[k]=\begin{bmatrix}
    \mathbf{B}[k] & 0 & 0 &\cdots & 0\\
    \mathbf{A}[k]\mathbf{B}[k] & \mathbf{B}[k] & 0 & \cdots & 0\\
    \mathbf{A}^2[k]\mathbf{B}[k] & \mathbf{A}[k]\mathbf{B}[k] & \mathbf{B}[k] & \cdots & 0\\
    \vdots & \vdots & \vdots & &\vdots \\
    \mathbf{A}^{N_\tau-1}[k]\mathbf{B}[k] & \mathbf{A}^{N_\tau-2}[k]\mathbf{B}[k] & \mathbf{A}^{N_\tau-3}[k]\mathbf{B}[k] & \cdots & \mathbf{B}[k]
    \end{bmatrix}
    $,
    }\\
    \mathbf{x}[k]=\begin{bmatrix}
    \rho_1[k]&
    \cdots&
    \rho_N[k]
    \end{bmatrix}
    ^\mathsf{T}
    \in \mathbb{R}^{N\times1},\\
    \mathbf{U}[k]=\begin{bmatrix}
    \mathbf{s}^\mathsf{T}[k]&
    \cdots&
    \mathbf{s}^\mathsf{T}[k+N_\tau-1]\\
    \end{bmatrix}
    ^\mathsf{T}
    \in \mathbb{R}^{\left(N_\tau N_{out}\right)\times 1}.
\end{gather}
\end{subequations}

Now, we can rewrite the cost function \eqref{cost function} as
\begin{equation}
\mathrm{C}=
  {1\over 2}\left(\mathbf{U}[k]^\mathsf{T}\mathbf{U}[k]\right).
\end{equation}

By using the predictive traffic coordination model \eqref{iterative model}, we can also rewrite the constraint equations \eqref{constrain condition} as follows:
\begin{subequations}\label{constrains}
\begin{gather}
    \mathbf{U}[k]\geq\mathbf{0}_{\left(N_\tau N_{out}\right)\times 1}, \label{u_constrain}\\
    \mathbf{G}[k]\mathbf{x}[k]+\mathbf{H}[k]\mathbf{U}[k]\leq\mathbf{1}_{N_\tau\times 1}\otimes\mathbf{x}_{\mathrm{max}}, \label{xmax_constrain}\\
    \mathbf{G}[k]\mathbf{x}[k]+\mathbf{H}[k]\mathbf{U}[k]\geq\mathbf{0}_{\left(N_\tau N\right)\times 1}, \label{xmin_constrain}\\
    \left(\mathbf{I}_{N_\tau}\otimes\mathbf{1}_{1\times N_\mathrm{out}}\right) \mathbf{U}[k]=d_0\mathbf{1}_{N_\tau\times 1}. \label{equality constrain}
\end{gather}
\end{subequations}
where $\mathbf{1}_{N_\tau\times1}\in\mathbb{R}^{N_\tau\times 1}$ and $\mathbf{1}_{1\times N_\mathrm{out}}\in\mathbb{R}^{1\times N_\mathrm{out}}$ are the one-entry vectors, $\mathbf{0}_{\left(N_\tau N_{out}\right)\times 1}\in\mathbb{R}^{\left(N_\tau N_{out}\right)\times 1}$ and $\mathbf{0}_{\left(N_\tau N\right)\times 1}\in\mathbb{R}^{\left(N_\tau N\right)\times 1}$ are the zero-entry vectors, and $\mathbf{I}_{N_{\tau}}\in\mathbb{R}^{N_{\tau}\times N_{\tau}}$ is the identity matrix. Eq. \eqref{u_constrain} integrates feasibility conditions  \eqref{c1} and \eqref{c2}. Constraint equations \eqref{xmax_constrain}, \eqref{xmin_constrain}, and \eqref{equality constrain} are identical to \eqref{c3}, \eqref{c4}, and \eqref{c5} respectively.
\begin{theorem}\label{TheoremNon-negative}
If $u_i[k]\geq 0$ at every $i\in \mathcal{V}_{in}$, $v_j[k]\geq 0$ at every $j\in \mathcal{V}_{out}$, $\rho_i[k]$ is updated by dynamics \eqref{Equation: 1}, and $\rho_i[0]\geq0$ at every node $i\in \mathcal{V}_I$, then $\rho_i[k]\geq 0$ at every interior road element $i\in \mathcal{V}_I$ and every discrete time $k$.
\end{theorem}
\textbf{Proof:}
By applying the mass conservation law in \eqref{Equation: 1}, traffic network dynamics can be obtained via Eq.\eqref{traffic dynamics interior node} at every interior road $i\in \mathcal{V}_I$. Per Assumption \ref{assumption1}, $p_i[k]\in [0,1)$ at every $i\in \mathcal{V}_I$ and all discrete times $k$, which indicate that $\left(1-p_i[k]\right)> 0$ on the right-hand side of Eq.\eqref{traffic dynamics interior node}. Also, $q_{i,j}[k]$ is defined as a quantity in interval [0,1] at every discrete sampling time $k$. If $\rho_i[0]\geq 0$, $u_i[k]\geq 0$ at every $i\in \mathcal{V}_{in}$, and $v_j[k]\geq 0$ at every $j\in \mathcal{V}_{out}$, then the right-hand side of Eq. \eqref{traffic dynamics interior node} must be a non-negative quantity at every discrete sampling time $k$. This implies that $\rho_{i}[k]\geq0$ at every node $i\in \mathcal{V}_I$ and discrete time $k$. 

Per Theorem \ref{TheoremNon-negative}, $\rho_i[k]\geq 0$ at every road $i\in \mathcal{V}_I$ and discrete time $k$. Therefore, condition \eqref{xmin_constrain} is redundant, and conditions \eqref{u_constrain}, \eqref{xmax_constrain}, and \eqref{equality constrain} are sufficient to determine the feasible optimal boundary input $\mathbf{U}^*[k]$ by solving the following quadratic programming problem:
\begin{equation}
    \min
  {1\over 2}\left(\mathbf{U}[k]^\mathsf{T}\mathbf{U}[k]\right)
\end{equation}
subject to equality constraint \eqref{equality constrain} and inequality constraint
\begin{equation}\label{inequality constrain sum}
    \begin{bmatrix}
    -\mathbf{I}_{N_\tau N_{out}}\\
    \mathbf{H}[k]
    \end{bmatrix}\mathbf{U}[k]\leq
    \begin{bmatrix}
    \mathbf{0}_{\left(N_\tau N_{out}\right)\times 1}\\
    \mathbf{1}_{N_\tau\times 1}\otimes\mathbf{x}_{\mathrm{max}}-
    \mathbf{G}[k]\mathbf{x}[k]
    \end{bmatrix}
    .
\end{equation}
Note that 
\begin{equation}
    \mathbf{s}^*[k]=\begin{bmatrix}
    \mathbf{I}_{N_{out}}&\mathbf{0}_{N_{out}\times \left(\left(N_\tau-1\right)N_{out}\right)}
    \end{bmatrix}
    \mathbf{U}^*[k]
\end{equation}
is the optimal boundary control at discrete time $k$.

\section{Simulation Results}\label{section 6}

In this section, we present the simulation results of modeling and control in the example NOIR shown in Fig.\ref{fig:Philadelphia Center}. This particular NOIR consists of $259$ road elements of Center City Philadelphia, where the index numbers of the road elements are shown in Fig.\ref{fig:Philadelphia Center}. We process the map data generated by SUMO and obtain the corresponding graph $\mathcal{G}\left(\mathcal{V},\mathcal{E}\right)$. Node set $\mathcal{V}=\{1,\cdots,259\}$ can be expressed as $\mathcal{V}=\mathcal{V}_{in}\bigcup \mathcal{V}_{out}\bigcup \mathcal{V}_{I}$ and $\mathcal{V}_{in}=\{1,\cdots,20\}, \mathcal{V}_{out}=\{21,\cdots,42\}, \mathcal{V}_{I}=\{43,\cdots,259\}$.

We set the whole simulation time as $3000s$ and the sampling interval as $20s$, which implies that the traffic coordination is simulated for $150$ time steps. At every discrete time $k\in \{1,\cdots,150\}$ the outflow probability matrix $\mathbf{P}[k]$ and the fraction probability matrix $\mathbf{Q}[k]$ are randomly generated  to simulate the uncertainty of human driving intent roughly. For simulation, we choose $u_0=100$, which implies $100$ cars are permitted to cross the boundary of the NOIR shown in Fig.\ref{fig:Philadelphia Center} at every discrete time  $k$. For every road $i\in \mathcal{V}_I$,
\begin{equation}
    \rho_{i,max}=\frac{n_{i,lane}*l_{i}}{l_{veh}}
\end{equation}
assigns an upper bound for the number of cars at road element $i\in \mathcal{V}_I$, where  $l_{veh}=4.5m$ is considered the same for all road elements, $l_{i}$ is the length of road element $i\in \mathcal{V}_I$ in the Center City area, and $n_{i,lane}$ refers to the number of lanes at road element $i\in \mathcal{V}_I$. Meanwhile, the initial traffic density $\rho_{i}[0]$ is assigned randomly for every road element $i\in\mathcal{V}$.

\begin{table}[h]
    \vspace{1.5em}
    \centering
    \caption{Example road elements in NOIR}
    \begin{tabular}{|m{1cm}|m{1.5cm}|m{4cm}|}
    \hline
    \textbf{Type}&\textbf{Road Index}&\textbf{Name and Location}\\
    \hline     
    Inlet&8&Cherry St. (N20th-N19th)\\
    \hline
    Inlet & 17 & S11th St. (Locust- Walnut)\\
    \hline
    Outlet & 27 & N16th St. (Race-Vine )\\
    \hline
    Outlet & 35 & Sansom St. (N20th-S19th)\\
    \hline
    Interior & 68 & Filbert St. (N12th-N13th )\\
    \hline
    Interior & 119 & JFK Blvd (N16th-N17th)\\
    \hline
    \end{tabular}
    \label{tab:my_label}
\end{table}

\begin{figure}[h]
    \centering
    \includegraphics[width=0.48\textwidth]{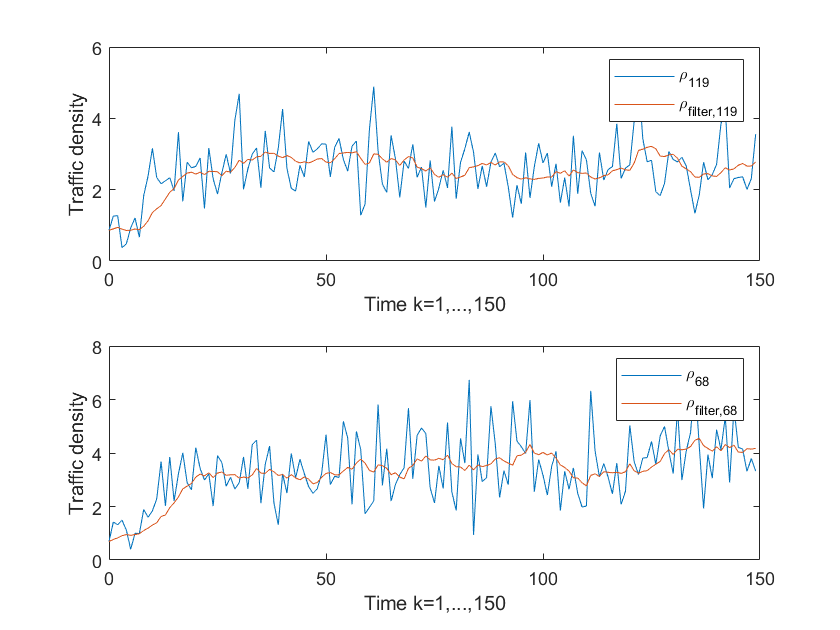}
    \caption{Variation trends of traffic densities at example interior road elements}
    \label{fig:interior density}
    \vspace{-2em}
\end{figure}

\begin{figure}[h]
\centering
\includegraphics[width=0.48\textwidth]{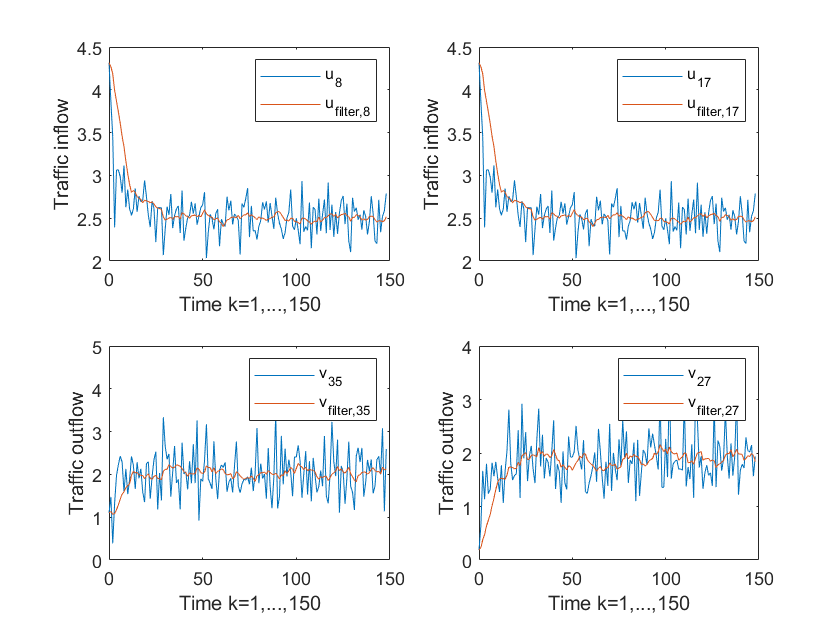}
\caption{External traffic flows at example inlet and outlet road elements}
\label{fig:example inlet and outlet flow}
\end{figure}

\begin{figure}[h]
\centering
\includegraphics[width=0.48\textwidth]{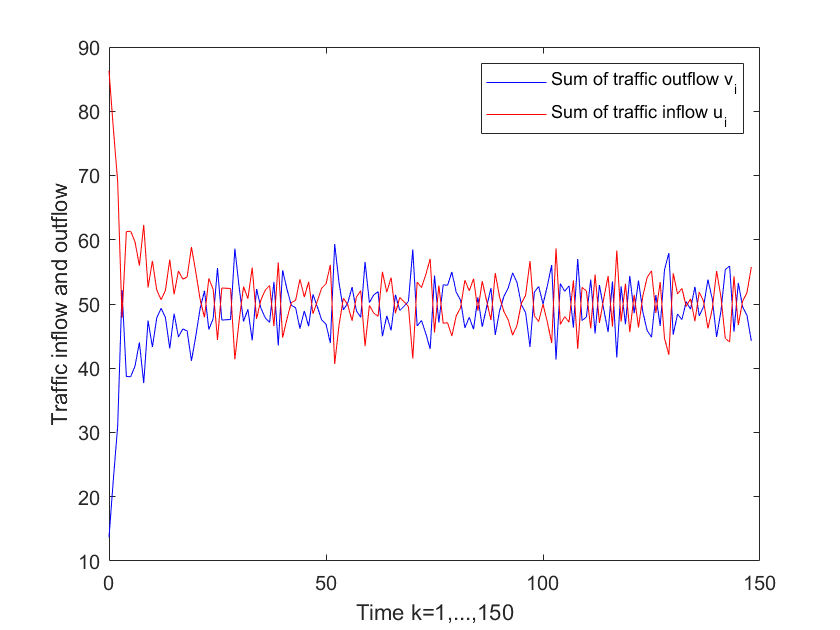}
\caption{External traffic inflow and outflow of the NOIR}
\label{fig:sum of external flow}
\end{figure}

We plot simulation results for two inlet boundary road elements $8,17\in \mathcal{V}_{in}$, two outlet boundary road elements $27,35\in \mathcal{V}_{out}$, and two interior road elements $68,119\in \mathcal{V}_{I}$. The locations of these six road elements are presented in Table \ref{tab:my_label}. 

Fig. \ref{fig:interior density} plots the density variations versus discrete sampling time $k$ at the example interior road elements. It can be observed that the traffic density of road elements 68 and 119 reaches the steady-state condition when the discrete sampling time $k>20$. 

Fig. \ref{fig:example inlet and outlet flow} illustrates the variation of the external traffic flow at inlet road elements $8,17\in \mathcal{V}_{in}$ and outlet road elements $27,35\in \mathcal{V}_{out}$. The variation trends at inlet and outlet road elements are similar: After a period of variation, the external traffic inflows and  the external traffic outflows reach the steady state condition. Fig. \ref{fig:sum of external flow} presents the net inflow and outflow of the NOIR versus discrete sampling time $k$. It could be observed that, after starting the simulation, the amount of traffic inflow decreases while the amount of traffic outflow increases symmetrically and after about 30 sampling times they converge to a stable state. We can formulate this variation trend as 
\vspace{-0.1em}
\begin{equation}
\sum_{i\in\mathcal{V}_{in}}u_i[k]=\sum_{i\in\mathcal{V}_{out}}v_i[k]\cong50,
\vspace{-0.4em}
\end{equation}
when $k>30$. 

\section{Conclusion}\label{section 7}
This paper introduce a conservation-based modeling method to learn the traffic network dynamics and alleviate the traffic congestion. We apply the mass conservation law to model traffic coordination by a time-varying stochastic process where the real map data is used to define the traffic network. We offer an MPC control to manage traffic congestion by controlling the inflow and outflow at the boundary of the NOIR. The simulation results demonstrate that our proposed modeling and control approach can manage the traffic congestion effectively through optimizing the traffic inflow and outflow across the boundary of the NOIR. In our future work, we plan to obtain the traffic dynamics based on real traffic data and control congestion through the boundary ramp meters and traffic signals, situated at road intersections.

\section*{acknowledgment}
The authors would like to acknowledge the Mechanical Engineering PhD fellowship provided to Xun Liu which was made possible by a generous gift from Dr. Yongping Gu and Fei Gu.

\bibliographystyle{asmems4}


\bibliography{reference}

\begin{thebibliography}{10}

\bibitem{muneera2018economic}
Muneera, C.~P., and Karuppanagounder, K., 2018.
\newblock ``Economic impact of traffic congestion- estimation and challenges''.
\newblock {\em European Transport / Trasporti Europei, {\bf 68}}.

\bibitem{margaret2004impact}
O’Mahony, M., and Finlay, H., 2004.
\newblock ``Impact of traffic congestion on trade and strategies for
  mitigation''.
\newblock {\em Transportation Research Board, {\bf 1873}}, pp.~25--34.

\bibitem{liang2013road}
Ye, L., Hui, Y., and Yang, D., 2013.
\newblock ``Road traffic congestion measurement considering impacts on
  travelers''.
\newblock {\em Journal of Modern Transportation, {\bf 21}}, pp.~28--39.

\bibitem{annan2015traffic}
Annan, J., Mensah, J., and Boso, N., 2015.
\newblock ``Traffic congestion impact on energy consumption and workforce
  productivity:empirical evidence from a developing country''.
\newblock {\em Archives of Business Research, {\bf 3}}, pp.~40--54.

\bibitem{chin2011impact}
Chin, H., and Rahman, M., 2011.
\newblock ``An impact evaluation of traffic congestion on ecology''.
\newblock {\em Planning Studies and Practice, {\bf 3}}, pp.~32--44.

\bibitem{xiao2014hierarchical}
Zhang, X., Onieva, E., Perallos, A., Osaba, E., and Lee, V.~C., 2014.
\newblock ``Hierarchical fuzzy rule-based system optimized with genetic
  algorithms for short term traffic congestion prediction''.
\newblock {\em Transportation Research Part C, {\bf 43}}, pp.~127--142.

\bibitem{toshio2017predicting}
Ito, T., and Kaneyasu, R., 2017.
\newblock ``Predicting traffic congestion using driver behavior''.
\newblock {\em Procedia Computer Science, {\bf 112}}, pp.~1288--1297.

\bibitem{zhao2019peak}
Huang, Z., Xia, J., Li, F., Li, Z., and Li, Q., 2019.
\newblock ``A peak traffic congestion prediction method based on bus driving
  time''.
\newblock {\em Entropy, {\bf 21}}, p.~709.

\bibitem{xiangjiekong2016urban}
Kong, X., Xu, Z., Shen, G., Wang, J., Yang, Q., and Zhang, B., 2016.
\newblock ``Urban traffic congestion estimation and prediction based on
  floating car trajectory data''.
\newblock {\em Future Generation Computer Systems, {\bf 61}}, pp.~97--107.

\bibitem{tettamanti2017nonlinear}
Tettamanti, T., Horváth, M.~T., and Varga, I., 2017.
\newblock ``Nonlinear traffic modeling for urban road network and related
  robust state estimation''.
\newblock In 9th European Nonlinear Dynamics Conference (ENOC 2017), EUROMECH,
  p.~247.

\bibitem{geroliminis2008existence}
Geroliminis, N., and Daganzo, C.~F., 2008.
\newblock ``Existence of urban-scale macroscopic fundamental diagrams: Some
  experimental findings''.
\newblock {\em Transportation Research Part B: Methodological, {\bf 42}}(9),
  pp.~759 -- 770.

\bibitem{xu2013traffic}
Xu, F., He, Z., Sha, Z., Sun, W., and Zhuang, L., 2013.
\newblock ``Traffic state evaluation based on macroscopic fundamental diagram
  of urban road network''.
\newblock {\em Procedia Social and Behavioral Sciences, {\bf 96}},
  pp.~480--489.

\bibitem{sirmatel2017integration}
Sirmatel, I.~I., and Geroliminis, N., 2017.
\newblock ``Integration of perimeter control and route guidance in large-scale
  urban networks via model predictive control''.
\newblock In Transportation Research Board 96th Annual Meeting, Transportation
  Research Board, p.~13p.

\bibitem{li2019model-free}
Li, Z., Jin, S., Xu, C., and Li, J., 2019.
\newblock ``Model-free adaptive predictive control for an urban road traffic
  network via perimeter control''.
\newblock {\em IEEE Access, {\bf 7}}, 11, pp.~172489--172495.

\bibitem{yang2017fundamental}
Yang, L., Yin, S., Han, K., Haddadc, J., and Hu, M., 2017.
\newblock ``Fundamental diagrams of airport surface traffic: Models and
  applications''.
\newblock {\em Transportation Research Part B: Methodological, {\bf 106}},
  pp.~29--51.

\bibitem{shao2018distributed}
Shao, P., Wang, L., Qian, W., Wang, Q.-G., and Yang, X.-H., 2018.
\newblock ``A distributed traffic control strategy based on cell-transmission
  model''.
\newblock {\em IEEE Access, {\bf 6}}, pp.~10771--10778.

\bibitem{munoz2003traffic}
Munoz, L., Sun, X., Horowitz, R., and Alvarez-Icaza, L., 2003.
\newblock ``Traffic density estimation with the cell transmission model''.
\newblock Vol.~5, pp.~3750 -- 3755.

\bibitem{yin2017offblock}
Yin, S., Yang, L., and Han, K., 2017.
\newblock ``Off-block flow optimisation based on cell transmission model''.
\newblock DASC.

\bibitem{feldman2002cell}
Feldman, O., and Maher, M., 2002.
\newblock ``A cell transmission model applied to the optimisation of traffic
  signals''.

\bibitem{greguric2020application}
Gregurić, M., Vujić, M., Alexopoulos, C., and Miletić, M., 2020.
\newblock ``Application of deep reinforcement learning in traffic signal
  control: An overview and impact of open traffic data''.
\newblock {\em Applied Sciences, {\bf 10}}, 06, p.~4011.

\bibitem{lin2018efficient}
Lin, Y., Dai, X., Li, L., and Wang, F.-Y., 2018.
\newblock An efficient deep reinforcement learning model for urban traffic
  control.

\bibitem{abdulhai2003reinforcementlearning}
Abdulhai, B., Pringle, R., and Karakoulas, G., 2003.
\newblock ``Reinforcement learning for true adaptive traffic signal control''.
\newblock {\em Journal of Transportation Engineering, {\bf 129}}, 05.

\bibitem{mannion2016anexperimental}
Mannion, P., Duggan, J., and Howley, E., 2016.
\newblock {\em An Experimental Review of Reinforcement Learning Algorithms for
  Adaptive Traffic Signal Control}.
\newblock 05, pp.~47--66.

\bibitem{prashanth2011reinforcementlearning}
L.A., P., and Bhatnagar, S., 2011.
\newblock ``Reinforcement learning with function approximation for traffic
  signal control''.
\newblock {\em Intelligent Transportation Systems, IEEE Transactions on, {\bf
  12}}, 07, pp.~412 -- 421.

\bibitem{rastgoftar2019integrative}
Rastgoftar, H., and Atkins, E., 2019.
\newblock An integrative data-driven physics-inspired approach to traffic
  congestion control.

\bibitem{rastgoftar2020resilient}
Rastgoftar, H., and Girard, A., 2020.
\newblock ``Resilient physics-based traffic congestion control''.
\newblock pp.~4120--4125.

\bibitem{rastgoftar2021physicsbased}
Rastgoftar, H., and Jeannin, J.-B., 2021.
\newblock A physics-based finite-state abstraction for traffic congestion
  control.

\bibitem{lin2011fast}
Lin, S., De~Schutter, B., Xi, Y., and Hellendoorn, H., 2011.
\newblock ``Fast model predictive control for urban road networks via milp''.
\newblock {\em Intelligent Transportation Systems, IEEE Transactions on, {\bf
  12}}, 10, pp.~846 -- 856.

\end{thebibliography}

\end{document}